# Triboelectric charge transfer theory driven by interfacial thermoelectric effect

**Authors:** Ji-Ho Mun[1,†], Eui-Cheol Shin[1,†], Jaeuk Seo[1], and Yong-Hyun Kim[1,2,*]

**Affiliations:**

[1]Department of Physics, Korea Advanced Institute of Science and Technology (KAIST), Daejeon 34141, Republic of Korea.

[2]School of Physics, Institute of Science, Suranaree University of Technology, Nakhon Ratchasima 30000, Thailand.

*Correspondence to: yong.hyun.kim@kaist.ac.kr

[†]These authors contributed equally.

**Abstract**

Despite extensive study and the practical significance of friction-driven static electricity, a quantitative triboelectric charge transfer theory has yet to be established. Here, we elucidate the detailed dynamics of triboelectric charge transfer driven by interfacial thermoelectric bias maintaining a steady state at the interface. We demonstrate that triboelectric charge exists in a delta-like distribution at a steady state. We suggest that the transferred triboelectric charge is dictated by half of the difference between thermoelectrically induced surface charges. Moreover, we quantitatively discuss electrostatic adhesion and static discharge between the transferred charges, which we may experience every day, including the role of surface charge inhomogeneity. Our findings may have significant implications for applications ranging from static electricity phenomena to advanced energy harvesting devices.

## Ⅰ. INTRODUCTION

Triboelectric charging occurs when two surfaces acquire electric charge through the energy dissipation process such as contact or friction. This phenomenon has productive applications such as xerography [1] and energy harvesting devices [2], but also detrimental onset such as sparks in microelectronics [3,4] and particle agglomeration in industrial polymers [5] or pharmaceuticals [6]. Moreover, triboelectric charging plays a key role in the solvent-free self-assembly of particles, organizing particles into ordered arrays without liquid media [7-10]. Recently, it has been reported that triboelectric charging has even an intimate influence on daily coffee flavor via granular agglomeration during the grinding of roast beans [11].

Despite its ubiquity, many researchers have struggled to understand triboelectric charging for several centuries. Historically, triboelectricity has been characterized using a triboelectric series, which empirically indicates the positive and negative charge between two materials [12]. However, the triboelectric series simply gives the qualitative charging trend, not the amount of charge transfer. Furthermore, the triboelectric series exhibits a low reproducibility and fails to account for several phenomena such as same-material charging [13-15] and triboelectric cycles [16,17]. To address these limitations, researchers have proposed various theoretical frameworks, including explanations based on (effective) work function difference [18-27], strain-driven polarization [28-31], and ion/material transfer [32-39]. Even, several studies have highlighted the role of water in triboelectric charging [11,40-44]. Nevertheless, while these arguments mainly focus on the underlying driving mechanisms of tribocharging, a quantitative prediction of charge transfer remains elusive.

Quantitative charge transfer models of triboelectricity have been proposed continuously, but with limitations. One of the widely proposed charge transfer models is the condenser (or capacitor) model, which treats the difference of either the work function or the contact potential as a potential difference across a capacitor [45,46]. It cannot explain the variations of charge transfer induced by several factors such as surface asperities and trap states. Thus, the model is usually modified by a correction with experimental parameters, but even the modified models provide limited explanations. Beyond the condenser model, additional models have been proposed, but they still rely on experimental fitting parameters [45,46]. Recently, a quantitative model of triboelectric charge transfer without empirical parameters based on flexoelectricity has been reported [47]. However, this model is still limited to the metal-semiconductor interface, making it difficult to generalize to tribocharging with insulators. Thus, a more comprehensive theory, explicitly addressing the magnitude of triboelectric charge transfer, is required.

Recently, Shin and co-workers proposed an exactly solvable triboelectric mechanism based on the interfacial thermoelectric effect, providing physical parameters to quantitatively explain the triboelectric series [48]. Note that Henry and Harper suggested that the thermoelectric effect may play a role in triboelectricity [22,49]. The newly revisited mechanism reveals that heat generated between two semi-infinite materials induces a charge redistribution via the thermoelectric effect under open circuit conditions. This thermally induced charge redistribution leads to a net charge accumulation at the surface. By quantitatively defining the triboelectric factor $\xi = S/\sqrt{\rho c k}$, where $S$ is Seebeck coefficient, $\rho$ is density, $c$ is specific heat, and $k$ is thermal conductivity, Shin and co-workers successfully estimated triboelectric charging polarities and quantified the thermoelectrically induced surface charge density. This surface charge density also exhibited saturation behavior with quantitative resemblance for experimental observations [48,50,51]. Moreover, theoretical concepts such as temperature profile and triboelectric factor have been gradually adopted in the triboelectric community [11,52-57]. Recently, Matsukawa and co-workers investigated the decomposition of molybdenum dithiocarbamate in nanoparticles and the formation of a $MoS_2$-containing tribofilm upon contact and release with diamond-like carbon and nanoparticles, and particularly explained triboemission properties based on the difference in the triboelectric factor of rutile (−0.154) and anatase (−0.102) $TiO_2$ nanoparticles [57]. Even if the triboelectric factor provides a quantitative guiding principle to estimate the charging polarity of materials, it has not settled the quantification of the charge transfer to establish thermal equilibrium in this process.

Here, we present a theoretical quantification method of triboelectric charge transfer driven by the interfacial thermoelectric effect, which has not been completely solved in our preceding study. Based on the steady state at the interface, we could elucidate that thermoelectric-driven triboelectric charge transfer is determined by half of each thermally induced surface charge as we discussed in previous research. The transferred charge exhibits saturation behavior over time and predictable behavior in triboelectric series. In addition, we discuss that static phenomena such as adhesion and discharge after tribocharging can be improved through surface charge inhomogeneity.

Integrating these insights, we develop a comprehensive theory of triboelectric charge transfer driven by thermoelectric effect in this work.

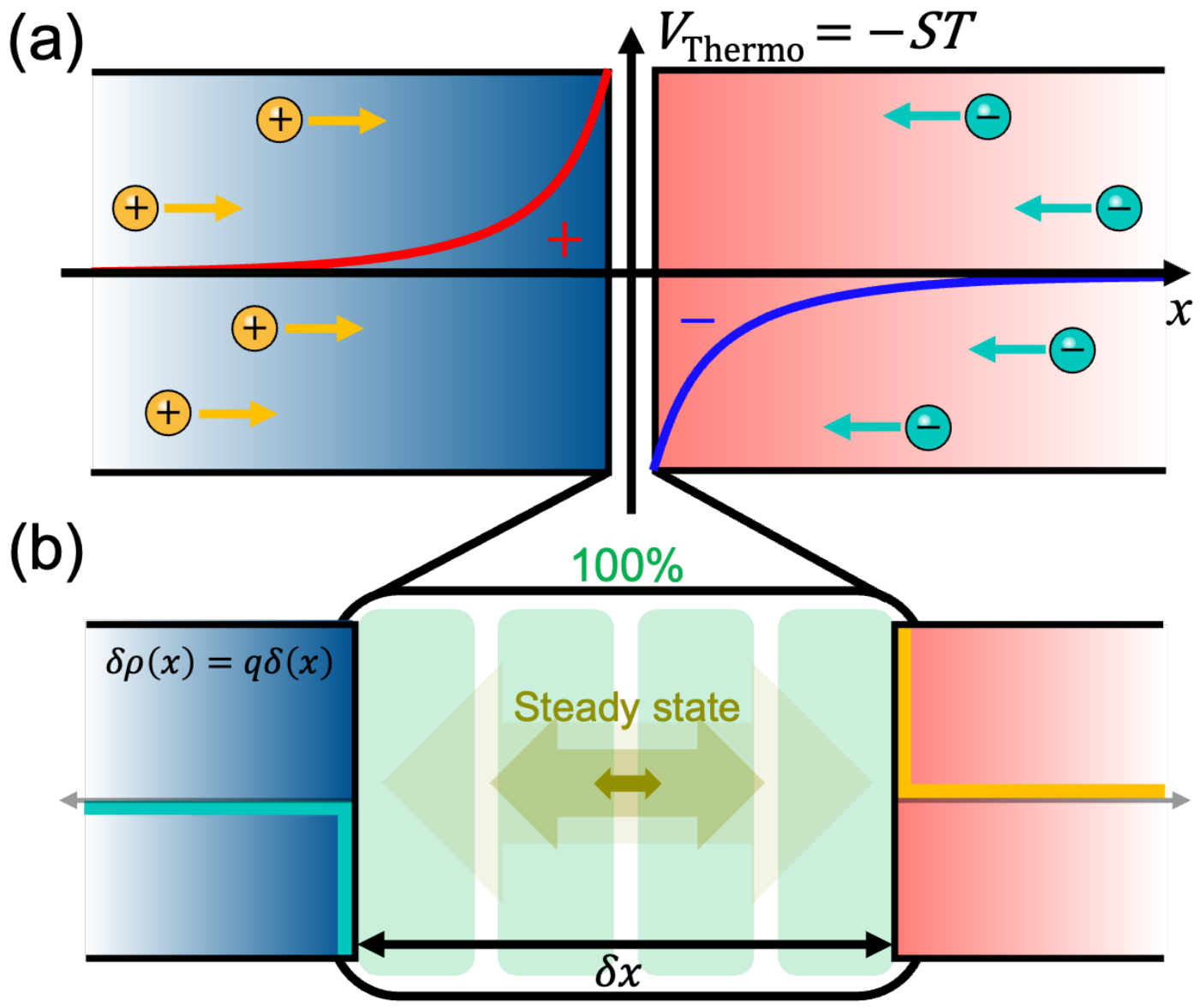

FIG. 1. Schematics of triboelectric charge transfer driven by thermoelectricity. (a) Voltage profile and charge accumulation caused by the thermoelectric effect. Charge carriers are redistributed due to the temperature gradient and develop a substantial voltage difference at the interface. (b) Steady state of triboelectric charging. The voltage difference in an infinitesimal gap $\delta x$ allows carriers to be transferred. Like the highly localized nature of charge diffusion in triboelectric systems, the transferred charge could be represented by Dirac delta distribution. Once triboelectric charge transfer is completed, a steady state at the interface is reached.

## Ⅱ. MAIN TEXT

Figure 1(a) shows that the thermodynamic driving force from thermoelectricity, i.e., a voltage variation is generated within each material. Considering the complexity and diversity of friction at interfaces, we simplified these complex phenomena by representing them as frictional heat. Microscopically, friction between two semi-infinite materials occurs when an external force causes surface atoms to cross energy barriers in the weakly interacting regime, which can be regarded as the breaking of van der Waals bonds. Then, frictional heat is generated via a bond-breaking process corresponding to stationary heat pulses $\dot{Q}(x,t) = Q_0\delta(x)\sum_n^N \delta(t - n\tau)$ at the interface, where $Q_0$ denotes the energy of the frictional heat produced per pulse and $\tau$ is the interval between pulses. Estimating the van der Waals energy dissipation in a typical sliding speed of 10 cm/s, we set $Q_0$ as 0.01 J/m$^2$ per $\tau$ = 1 ns with a friction coefficient of $\mu$ = 0.1 [48]. Since the use of a half-and-half heat partition doesn't change the underlying physics and is insensitive to interfacial thermal conductance, the temperature profile from the one-dimensional heat equation is analytically derived as follows

$$T(x,t) = \frac{1}{2}\frac{Q_0/\tau}{\sqrt{\pi\rho ck}}\sqrt{t}\mathcal{E}_{3/2}\left[\frac{x^2}{4\alpha t}\right], \qquad (\text{as } t \gg \tau) \tag{1}$$

where $\mathcal{E}_m(z) = \int_1^\infty e^{-zu}/u^m\, du$ is exponential integral function and $\alpha$ is thermal diffusivity. The thermoelectric response is represented by $J = \sigma(E - S\nabla T)$ where $J$ is the current density, $\sigma$ is the electrical conductivity, $E$ is the $E$-field, and $T$ is the temperature distribution [48,58]. Considering that the dynamic of electrons is in femtoseconds [59], an instantaneous equilibrium or open circuit condition ($J$ = 0) for temperature provides the charge distribution $\rho_0 = \varepsilon S\nabla^2 T$ where $\varepsilon$ is the dielectric constant. This consideration allows the current density $J$ to be treated in terms of the potential $V$ with a steady-state condition, where $V = -ST$. The potential difference $\Delta V$ at the interface $x$ = 0 of two contacting materials is

$$\begin{aligned}\Delta V &= V_\mathrm{L} - V_\mathrm{R} = S_\mathrm{R} T_\mathrm{R}(0,t) - S_\mathrm{L} T_\mathrm{L}(0,t)\\ &= \frac{Q_0}{\tau}\sqrt{t/\pi}\left[\frac{S_\mathrm{R}}{\sqrt{\rho_\mathrm{R} c_\mathrm{R} k_\mathrm{R}}} - \frac{S_\mathrm{L}}{\sqrt{\rho_\mathrm{L} c_\mathrm{L} k_\mathrm{L}}}\right],\end{aligned} \tag{2}$$

where subscripts L and R correspond to $x < 0$ and $x > 0$. This potential gap at the interface acts as a quantitative indicator of triboelectric hierarchy or polarity via the triboelectric factor $\xi$ [48]. Currently, the material properties within this factor have used bulk properties as the first-order approximation to establish triboelectric series or polarity, but a more precise description of the thermoelectric-driven triboelectricity theory at the interface could be achieved if material properties that accurately capture interfacial variations were used. The interfacial thermoelectric response generates a thermally accumulated surface charge $\sigma_\mathrm{surface} = -\varepsilon S Q_0/(2k\tau)$ for the input heat $Q_0$. $\sigma_\mathrm{surface}$ should be transferred due to the interfacial voltage gap, which has not been quantitatively discussed in our previous theory.

Figure 1(b) represents that the transferred triboelectric charge $q$ has the distribution of delta-like function, $\delta\rho(x) = q\delta(x)$, once a steady state at the interface is reached. A highly localized charge distribution on the surface is frequently observed in triboelectric systems [60-62]. Since the bulk thermoelectric response is already satisfied as $E = S\nabla T$, the transferred charge concentrates at the surface, forming the delta-like distribution. Moreover, this charge distribution is unique, as any deviation tends to decay over time (See Supplementary Note 1 for details). Based on this localized charge distribution, we could investigate the dynamics of the transferred charge by analyzing the interfacial electric field.

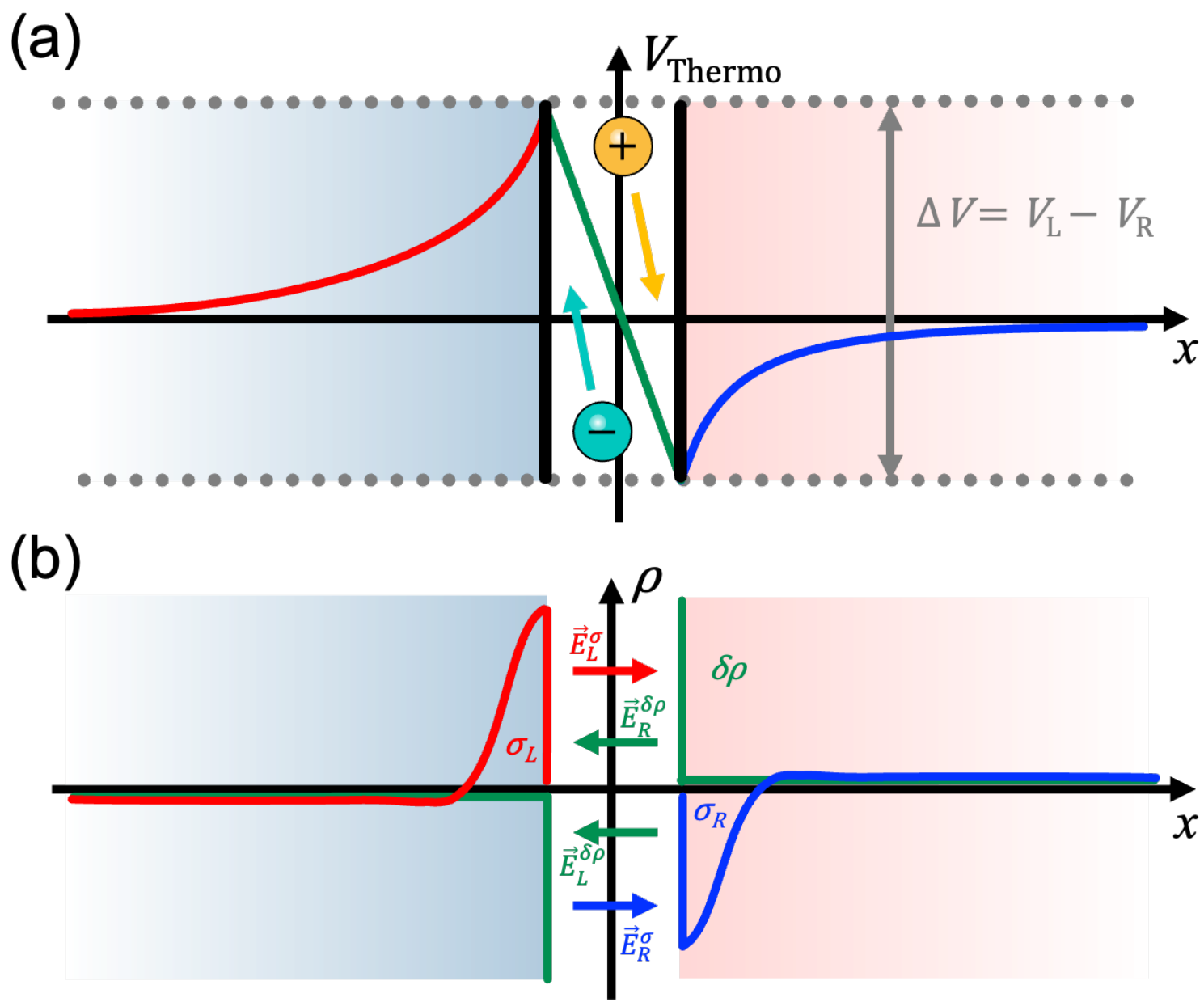

FIG. 2. Schematics of the charging mechanism at the interface. (a) The interfacial voltage gap induced by thermally induced surface charge. Charges accumulated by the thermoelectric voltage profile induce interfacial voltage gap $\Delta V$ and thus can be transferred between two surfaces. (b) The interfacial $E$-field generated by thermally induced surface charge and the transferred charge. The surface charges $\sigma_L$ and $\sigma_R$, which are driven by the thermoelectric voltage profile, generate $E$-fields of $\vec{E}_L^{\sigma}$ and $\vec{E}_R^{\sigma}$, respectively. The transferred charge distribution $\delta\rho$ gives $\vec{E}_L^{\delta\rho}$ and $\vec{E}_R^{\delta\rho}$. For the steady state at the interface, the total $E$-field should be zero.

In Fig. 2(a), the thermoelectric potential difference $\Delta V$ enables charge transfer in the direction of interfacial voltage offset. A reverse bias formed by the transferred charge $q$ in the unit of μC/m$^2$ leads to a steady state at the interface. Charge transfer occurs until the transferred charges compensate thermoelectric-induced charges corresponding to $\Delta V$. Figure 2(b) illustrates the interfacial $E$-field generated by thermally induced surface charges ($\sigma_{L,R}$) and the transferred charges ($\delta\rho$). For the $\delta$-type charge distribution, the transferred charges induce the $E$-field in the infinitesimal gap only. Thus, the interfacial $E$-field $\vec{E}_{\text{int}}$ is formulated as follows,

$$\vec{E}_{\text{int}} = \vec{E}_{\text{L}}^{\delta\rho} + \vec{E}_{\text{R}}^{\delta\rho} + \vec{E}_{\text{L}}^{\sigma} + \vec{E}_{\text{R}}^{\sigma} \tag{3a}$$

$$= -\frac{q}{\varepsilon_0} + \left(-\frac{\varepsilon_{\text{r,L}}}{2}\frac{S_{\text{L}}}{k_{\text{L}}} + \frac{\varepsilon_{\text{r,R}}}{2}\frac{S_{\text{R}}}{k_{\text{R}}}\right)\frac{Q_0}{2\tau}, \tag{3b}$$

where $\varepsilon_0$ and $\varepsilon_r$ are vacuum and relative permittivity, respectively. The first two terms $\vec{E}_{\text{L,R}}^{\delta\rho}$ in Eq. 3(a) represent the $E$-field induced by the transferred charges, which is the same as the $E$-field between two infinite plates. The last two terms are the $E$-field caused by the thermally induced surface charges. Since $\vec{E}_{\text{int}}$ is zero once the steady state at the interface is reached, the transferred charge $q$ in Eq. 3(b) is as follows,

$$q = \left(-\frac{\varepsilon_{\text{L}}}{2}\frac{S_{\text{L}}}{k_{\text{L}}} + \frac{\varepsilon_{\text{R}}}{2}\frac{S_{\text{R}}}{k_{\text{R}}}\right)\frac{Q_0}{2\tau} = \frac{\sigma_{\text{L}} - \sigma_{\text{R}}}{2}, \tag{4}$$

where $\varepsilon_{L,R}$ and $\sigma_{L,R}$ are dielectric constants and the $\sigma_{\text{surface}}$ of the left and right materials, respectively. Equation 4 shows that the transferred charge $q$ is governed by half of the difference between thermally induced surface charge densities in each material. In Eq. 1, the temperature profile generated by the frictional heat follows the function of $\mathcal{E}_{3/2}[x^2/4\alpha t]$. As time goes to infinity, the temperature profile near the interface can be approximated linearly as $T(x,t) \approx \frac{Q_0/\tau}{\sqrt{\pi\rho c k}}\sqrt{t} - \frac{Q_0/\tau}{2k}x$, corresponding to estimating the charge distribution at a steady-state temperature [63]. In this linear approximation, the thermoelectric response for charge carriers takes the form of the delta-like function at the surface. Consequently, after sufficient time passes, thermally accumulated surface charges directly influence the amount of the transferred charge.

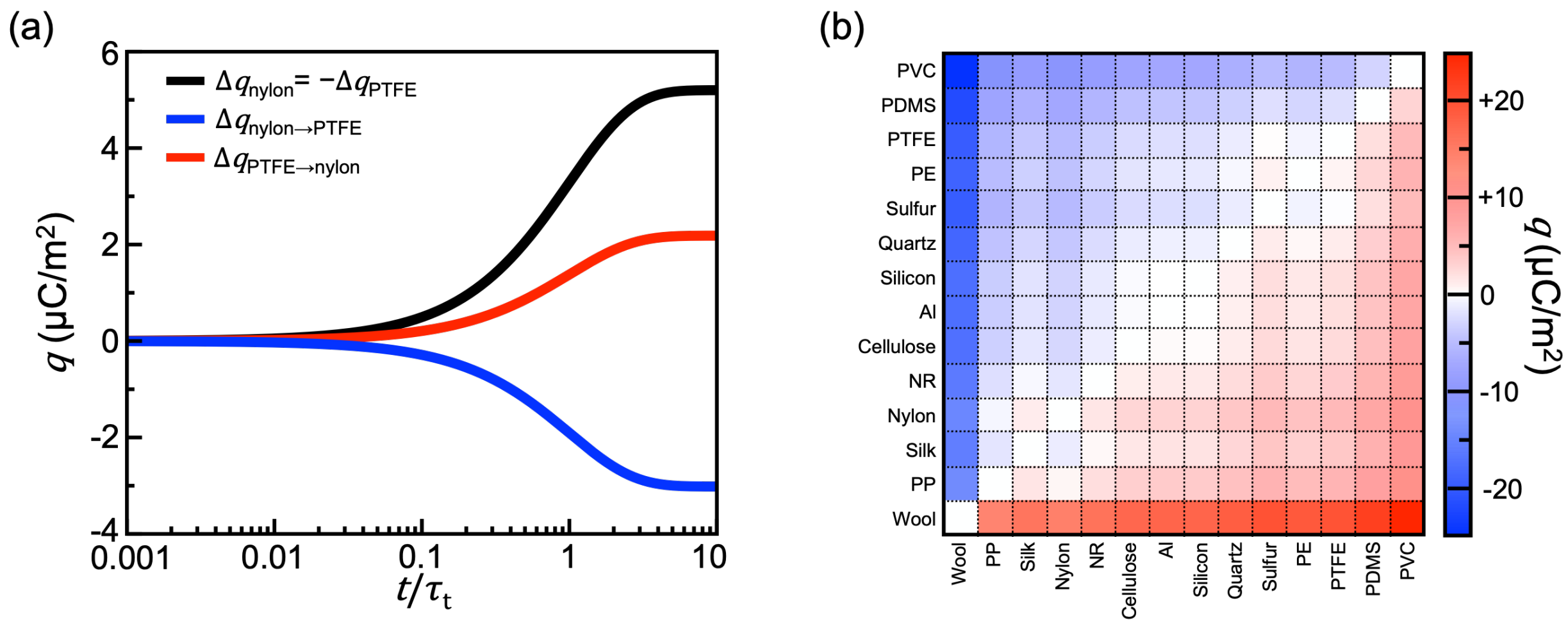

FIG. 3. Charging dynamics as a function of time and triboelectric pairs. (a) Charging dynamics of PTFE-nylon. $\tau_t$ is a time constant defined by $\tau_t^{-1} = \sigma_{\text{int}}/\varepsilon_0$. The black line represents the net transferred charge between the PTFE-nylon pair. The nylon and PTFE are positively and negatively charged, respectively. The blue and red lines are charge transfers from nylon to PTFE and vice versa. (b) Comparison of charging between materials in triboelectric series. All values correspond to the transferred charges of $y$-axis materials rubbed against $x$-axis materials. The red and blue colors are positively and negatively charged, respectively.

Figure 3 shows the charging dynamics as a function of time and triboelectric pairs when subjected to frictional heat. To calculate the transferred charge $q$ between triboelectric pairs, we obtain thermally induced surface charges for the triboelectric pairs using theoretical Seebeck coefficients, experimental thermal conductivities [48], and the experimental dielectric constants [64-75]. The theoretical Seebeck coefficient was estimated by comparing the charge neutrality energy from electronic structures to the hydrogen redox potential of −4.44 eV, which represents the probable Fermi level for materials in inevitable moisture [48,76].

Figure 3(a) shows the charging behavior of the PTFE-nylon pair, which is the most reproducible combination in the experiment. The transferred charge $q$ presents the saturation behavior with a time constant $\tau_t$ estimated to be roughly 10 μs (See Supplementary Note 2 for details). This implies that triboelectric charging occurs in very short bursts. Because the PTFE and nylon have negative and positive Seebeck coefficients; $S_{\text{PTFE}} = -12.3$ mV/K and $S_{\text{nylon}} = +11.1$ mV/K, each PTFE and nylon develops positive and negative $\sigma_{\text{surface}}$, respectively. Following Eq.

S3, each path in Fig. 3(a) means the charge transfer between a certain $\sigma_{\text{surface}}$ and a neutral one (or superconductor). For example, the $\sigma_{\text{surface}}$ of PTFE with a positive sign accepts an electron from a counterpart neutral material. The nylon donates electrons to the counterpart system. The $\sigma_{\text{surface}}$ of PTFE and nylon is calculated at +4.4 and −6.0 μC/m$^2$, which leads to $q$ = 5.2 μC/m$^2$ in Fig. 3(a). Notably, the theoretical estimation is quantitatively comparable to the experimental values of ~3 μC/m$^2$ [51].

Finally, we calculated the transferred charges of $y$-axis materials rubbed against $x$-axis materials in triboelectric series using Eq. 4, as shown in Fig. 3(b). The order of triboelectric series comes from the previous study with a remarkable similarity between theoretical and experimental triboelectric series [48]. The more triboelectric pairs are separated, the more charge is usually transferred, which can be up to about 25 μC/m$^2$ for wool-PVC. Several pairs, such as nylon-silk, PE-sulfur, and PTFE-sulfur show reverse polarity in triboelectric series because nylon, PE, and PTFE exhibit larger $\sigma_{\text{surface}}$s than silk and sulfur. For the nylon-silk pair, it has been reported that nylon is positively charged, and silk is negatively charged [77,78]. For PE-sulfur and PTFE-sulfur pairs, the dielectric constant of solid sulfur is used to be 3.69 [73]. However, it can be from 1.5 to 1.8 in a compendium [79], which can change the sign and amount of the transferred charge. Moreover, the charging behavior of PTFE-Al and nylon-Al, i.e., insulator-metal is solely determined by the $\sigma_{\text{surface}}$ of insulators. The $S_{\text{Al}}$ = −2.5 μV/K is almost negligible, which leads to an insignificant magnitude of $\sigma_{\text{surface,Al}}$. As a result, Eq. 4 is approximated to $q = \sigma_{\text{surface,ins}}/2$. Additionally, the quantification of charge transfer is indirectly applicable to the measurement of the Seebeck coefficient of insulators via the formula $\sigma_{\text{surface}} = -\varepsilon S Q_0/(2k\tau)$, which may be considered the most challenging issue in thermoelectric research.

## Ⅲ. DISCUSSION

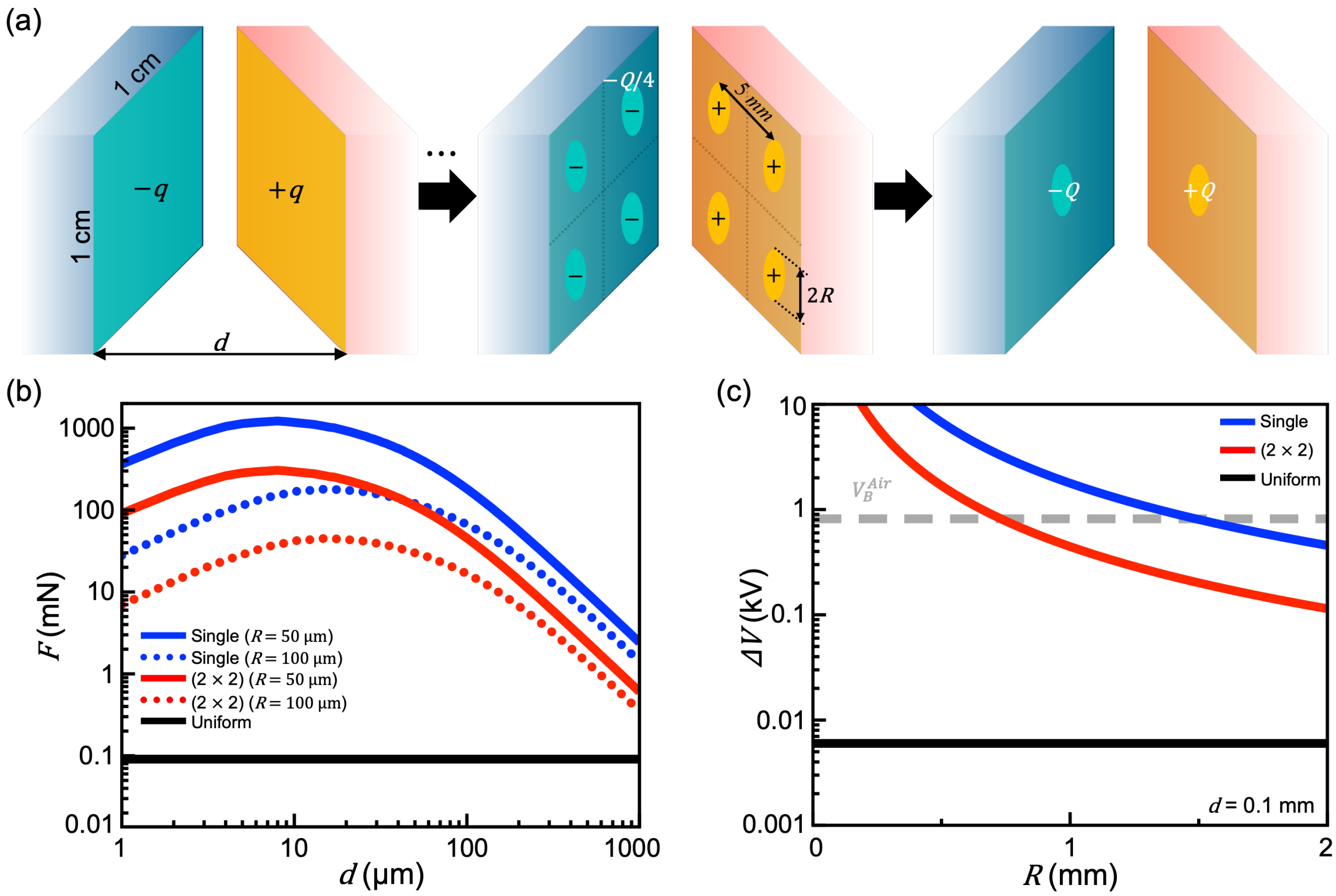

FIG. 4. Electrostatic force and discharge after triboelectric charging. (a) Schematic representation of surface charge localization. To test the inhomogeneity of the triboelectric surface charge, we evaluate the electrostatic force and discharge on two cases of localized triboelectric charges: single and (2 × 2) pattern of circular charges. A uniform surface charge density $q$, which is equal to the transferred charge, on an area $A$ = 1 cm$^2$ gives a total charge $Q = qA$. Each of the four circles in the (2 × 2) pattern therefore carries $Q/4$. (b) Electrostatic force $F$ exerted by triboelectric charges for the PTFE-nylon pair. The black line represents the force by the uniform charge. The blue and red lines show the force of the single and (2 × 2) charges, respectively. The solid and dotted lines correspond to circular-charge radii of 50 and 100 μm, respectively. (c) Potential difference $\Delta V$ produced by triboelectric charges for the PTFE-nylon pair. The blue, red, and black solid lines show the potential difference of the single, (2 × 2), and uniform charges, respectively. The grey dashed line marks the breakdown voltage of air given by Paschen's law. The gap distance between the two materials is 0.1 mm.

After triboelectric charging, electrostatic phenomena such as adhesion and discharge typically arise from the transferred triboelectric charges. To investigate these phenomena in detail, we examine how triboelectric charges influence the magnitude of electrostatic forces and the conditions for discharge. Although our model has considered that triboelectric surface charge is formed uniformly on flat surfaces, real material surfaces often exhibit an uneven distribution of charge due to surface morphology [80-82]. Since precisely analyzing realistic scenarios involving complex surface morphologies is challenging, we introduce a simple model with highly localized triboelectric surface charges, as shown in Fig. 4(a). Specifically, we assume that the charge is confined as the single and (2 × 2) patterns within a circular region of radius $R$ on contact area $A$ = 1 cm$^2$. The transferred charge $q$ on this area gives a total charge $Q = qA$, where the (2 × 2) charges have a charge of $Q/4$ per circle. We set a separation distance between circular charges as 5 mm in the (2 × 2) charges, where all circular charges are equally located within the plane. By examining these cases of localized charge, we can identify the impact of charge inhomogeneity on electrostatic forces and discharge conditions varying the gap distance $d$ and the radius $R$.

Figure 4(b) compares the electrostatic forces due to the transferred charges between the uniform charge and circular charges. In the uniform charge, the electrostatic force $F = q^2A/2\varepsilon_0$ for PTFE-nylon is about 0.15 mN with $q$ = 5.2 μC/m$^2$, which can lift light objects weighing 0.015 mg. On the other hand, using Eq. S4, the electrostatic force of the single and (2 × 2) charges can be larger than that for the uniform charge (See Supplementary Note 3 for calculation details). When the radius of localized charges decreases, the force can increase. In particular, the force in the specific distance has the local maximum. This implies that a stronger force is applied to prevent the charged materials from separating each other. Furthermore, since the thermally induced surface charge is proportional to the input heat $Q_0$, the thermoelectric-driven charge transfer could produce enormous force if additional heat sources like joule heating or photothermal heating are considered. Then, this force can be switched and its magnitude adjusted depending on the heat injection, enabling to apply electro-adhesion to move or secure objects.

We also analyze the maximum potential difference $\Delta V$ generated by the localized charges with the breakdown voltage from Paschen's law, which is an empirical relation that defines the breakdown voltage necessary to initiate a discharge in various gases. Since triboelectric charging is typically generated in the air, we adopt Paschen's law of air [83]. Also, we use 1 atm of pressure by assuming slow separation of materials. As the small gap distance $d$ between two materials fails to predict the breakdown voltage [84], we select $d$ = 0.1 mm where Paschen's law is valid [85]. In Fig. 4(c), the potential difference of the uniform charge for the PTFE-nylon pair, $\Delta V_{\mathrm{uniform}} = qd/\varepsilon_0$, doesn't reach the breakdown voltage of air, which means that triboelectric discharge cannot occur. On the other hand, the electric potential of the single and (2 × 2) charges with a radius smaller than a certain radius becomes larger than the breakdown voltage (See Supplementary Note 3 for calculation details). This implies that localized charges can be one of the factors leading to triboelectric discharge.

Our theory addresses triboelectric charging between semi-finite materials driven by interfacial thermoelectric effects. According to our analysis, no driving force for triboelectric charging exists between identical materials. However, we predict that triboelectric charging between identical materials is primarily governed by higher-order interfacial effects such as material size, surface roughness, and curvature. When these effects are considered, even identical materials may exhibit different interfacial temperature profiles, thus enabling tribocharging. Indeed, it has been reported that triboelectric charging is enhanced on bumpy surfaces [86-88] and even occurs between identical materials [89-91]. To explain this, there is ongoing research into how much the charge or electrical interaction changes with surface roughness, but it is not yet well understood [92-94]. Future work will delve into these higher-order interfacial effects and their impacts on frictional heat generation and charge-transfer behavior. A rigorous understanding of these higher-order effects would be expected to explain charge mosaics, where positive and negative charges alternately coexist on a single contact surface [80,95-97].

As we mentioned in the introduction, water affects tribocharging in several ways [11,40-44]. Thus, the influence of humidity requires systematic exploration. In our model, we consider humidity to play a role in establishing a global Fermi level, thereby influencing the theoretical Seebeck coefficient [48,76]. However, water may also serve as a lubricant to reduce energy dissipation or affect heat transfer through absorption and evaporation. Moreover, water can also function as the number of defects, changing the chemical potential or affecting the surface morphology. A rigorous study of how water influences heat generation and charge transfer, as well as the ultimate role it plays in thermoelectric-driven tribocharging, is thus indispensable.

By analyzing how higher-order interfacial effects influence thermoelectric-driven triboelectric charging dynamics, we seek to develop a predictive framework for optimizing energy harvesting in triboelectric nanogenerators through engineered micro/nanostructured surfaces and tunable thermal gradients. This enables self-regulating triboelectric nanogenerators with adaptive power output under varying conditions. Beyond energy harvesting, controlled charge modulation has potential applications in electro-adhesion for robotics, precision particle assembly in additive manufacturing, and electrostatic haptic interfaces.

## Ⅳ. CONCLUSION

In conclusion, thermoelectric-driven charge transfer offers a new way of achieving controllable triboelectric charge. We have confirmed that the steady-state condition at the interface induces the balance between the thermodynamic driving force by the thermoelectric response in bulk charge carriers and the $E$-field from the transferred charge. The triboelectric factor is a more fundamental figure-of-merit to determine absolute charging polarity, but the thermally induced surface charge is a good indicator for the resultant transferred charge. By systematically tuning variables such as density, specific heat, thermal conductivity, and thermopower, our findings can be leveraged to optimize triboelectric nanogenerators and enhance their output, demonstrating the practical applicability of our approach in energy harvesting. Nevertheless, our model currently has limitations, particularly in explaining triboelectric charging between identical materials and fully addressing environmental factors such as humidity and temperature variations. Future research should systematically explore these aspects to enhance the robustness and applicability of the theory. Finally, we propose that the validation of thermoelectric-driven charge transfer could be realized via the control of thermally induced surface charge using systematically controllable thermoelectric materials such as metal-organic framework and carbon nanotube sheets, or via naught charging between superconductors with zero Seebeck coefficient [12].

## ACKNOWLEDGMENTS

This work was supported by the Grand Challenge 30 program from the College of Natural Sciences, KAIST.

# Supplementary Information for

## Triboelectric charge transfer theory driven by interfacial thermoelectric effect

Ji-Ho Mun[1,†], Eui-Cheol Shin[1,†], Jaeuk Seo[1], and Yong-Hyun Kim[1,2,*]

[1]Department of Physics, Korea Advanced Institute of Science and Technology (KAIST), Daejeon 34141, Republic of Korea.

[2]School of Physics, Institute of Science, Suranaree University of Technology, Nakhon Ratchasima 30000, Thailand.

*Correspondence to: yong.hyun.kim@kaist.ac.kr

[†]These authors contributed equally.

**Supplementary Note 1**

To demonstrate the uniqueness of the charge distribution, we assume an arbitrary total charge density $\rho_{tot} = \rho_{sol} + \delta\rho$, where $\rho_{sol}$ is the solution of triboelectric charge, and $\delta\rho$ is a deviated charge density. As $\rho_{tot}$ is not a steady state, $\delta\rho$ must undergo time evolution. Therefore, the time derivative of the charge distribution in the continuity equation is the following

$$\frac{\partial \rho_{\text{tot}}}{\partial t} = -\frac{\partial}{\partial x} J_{\text{tot}} = -\frac{\partial}{\partial x}\left(J_{\text{sol}} + J_{\delta\rho}\right), \tag{S1}$$

where $J$ is the current density. Since $\rho_{sol}$ already satisfies the continuity equation, the terms of $\delta\rho$ in Eq. S1 are left. Using Ohm's law and Gauss's law, the time derivative of the deviated charge is following

$$\frac{\partial}{\partial t}\delta\rho = -\frac{\partial}{\partial x} J_{\delta\rho} = -\sigma \frac{\partial}{\partial x} E_{\delta\rho} = -\frac{\sigma}{\varepsilon}\delta\rho, \tag{S2}$$

where $\sigma$ is electrical conductivity and $\varepsilon$ is the dielectric constant. Thus, $\delta\rho$ in Eq. S2 shows an exponentially decaying behavior as time flows, and the delta-like solution at the steady state is unique.

**Supplementary Note 2**

Since the total current density at the interface is $J_{\mathrm{int}} = \sigma_{\mathrm{int}}\vec{E}_{\mathrm{int}}$ where $\sigma_{\mathrm{int}}$ is the interfacial electrical conductivity, Eq. 3(b) becomes,

$$J_{\mathrm{int}} = -\frac{\sigma_{\mathrm{int}}}{\varepsilon_0}q + \sigma_{\mathrm{int}}\left(-\frac{\varepsilon_{\mathrm{r,L}}}{2}\frac{S_{\mathrm{L}}}{k_{\mathrm{L}}} + \frac{\varepsilon_{\mathrm{r,R}}}{2}\frac{S_{\mathrm{R}}}{k_{\mathrm{R}}}\right)\frac{Q_0}{2\tau}. \tag{S2}$$

The determination of $\sigma_{\mathrm{int}}$ in various interfaces is practically challenging. Considering that the $J_{\mathrm{int}}$ of a typical triboelectric nanogenerator is sub-μA per tens of $\mathrm{cm}^2$, the $\sigma_{\mathrm{int}}$ can be conservatively estimated to be 1 μS/m [S1]. What is noteworthy here is that the steady state of $J_{\mathrm{int}}$ i.e., effectively naught current is not established with $q = 0$. Using the definition of $J_{\mathrm{int}} = \mathrm{d}q/\mathrm{d}t$, the following charge dynamics $q(t)$ arrives at

$$q(t) = \varepsilon_0\left(-\frac{S_{\mathrm{L}}}{k_{\mathrm{L}}} + \frac{S_{\mathrm{R}}}{k_{\mathrm{R}}}\right)\frac{Q_0}{2\tau}\left(1 - e^{-t/\tau_{\mathrm{t}}}\right) \tag{S3}$$

where $\tau_{\mathrm{t}}$ is a time constant defined by $\tau_{\mathrm{t}}^{-1} = \sigma_{\mathrm{int}}/\varepsilon_0$, and $\varepsilon_{\mathrm{r}}$ is the relative permittivity of each material. Substituting 1 μS/m for $\sigma_{\mathrm{int}}$ gives an instantaneous charging time of about 10 μs.

**Supplementary Note 3**

Figure 4(a) shows the single and (2 × 2) triboelectric surface charges. To calculate their electrostatic force, we set two circular charges with the radius *R*, whose centers are located on the origin and the position (*a*, *b*, *d*), respectively. Then, the electrostatic force between two circular charges is as follows

$$F_{circle} = \frac{Q^2 d}{4\pi\varepsilon_0(\pi R^2)^2}\int_0^R\int_0^{2\pi}\int_0^R\int_0^{2\pi} r_1 r_2[r_1^2 + r_2^2 + 2a^2 + d^2 - 2r_1(a\cos\theta + b\sin\theta) + 2r_2(a\cos\phi + b\sin\phi) - 2r_1 r_2\cos(\theta - \phi)]^{-3/2} dr_1 dr_2 d\theta\, d\phi, \quad \text{(S4)}$$

where $r_{1,2}$ is the radial coordinates in each circle, and $\theta$, $\phi$ are azimuth angles on each circle, respectively. *a*, *b* are the separation distances between circular charges, which use 5 mm in this work. By setting $a = b = 0$, the electrostatic force of the single charge can be calculated. On the other hand, the electrostatic force of the (2 × 2) charges is the sum of all forces between circular charges, substituting *Q* into *Q*/4 in Eq. S4. Except for forces between circular charges in the same plane, the electrostatic force of the (2 × 2) charges consists of the sum of four $F_{circle}$ with (*a*, *b*) = (0, 0), twelve $F_{circle}$ with (0, 5 mm), and four $F_{circle}$ with (5 mm, 5 mm).

For a circular charge with the radius *R* on the origin, the electric potential is following

$$V_{circle}(x, y, z) = \frac{Q}{\varepsilon_0 R^2}\int_0^R \frac{s\, ds}{\sqrt{\left(\sqrt{x^2 + y^2} + s\right)^2 + z^2}} K\left(\sqrt{\frac{4rs}{\left(\sqrt{x^2 + y^2} + s\right)^2 + z^2}}\right), \quad \text{(S5)}$$

where *K*(*x*) is the elliptic integral of the first kind. If the center of the circular charge is (*a*, *b*, *d*), its potential is substituting *x*, *y*, and *z* of Eq. S5 into $x - a$, $y - b$, and $z - d$, respectively. The potential of the single charge is the sum of the potential between two circular charges, whose centers are located on the origin and the position (0, 0, *d*), respectively. Then, we calculate the potential difference between the origin and the position (0, 0, *d*), which has the maximum value. For the (2 × 2) charges substituting *Q* into *Q*/4 in Eq. S5, the potential is the sum of the potential between eight circular charges with the spacing $a = b = 5$ mm. Like the case of the single charge, we calculate the maximum potential difference between the origin and the position (0, 0, *d*). No matter which center of the circular charge we choose, the potential difference is the same.